\documentclass{egpubl}
\usepackage{eurovis2026}

\SpecialIssuePaper         

\CGFccby

\usepackage[T1]{fontenc}
\usepackage{dfadobe}  

\usepackage{cite}  
\BibtexOrBiblatex
\electronicVersion
\PrintedOrElectronic
\ifpdf \usepackage[pdftex]{graphicx} \pdfcompresslevel=9
\else \usepackage[dvips]{graphicx} \fi

\usepackage{egweblnk}

\usepackage{trrracerLink}
\usepackage{externallink}
\usepackage{review}

\usepackage[dvipsnames]{xcolor}
\newcommand{\rev}[1]{{\textcolor{Black}{\normalsize {#1}}}}

\usepackage{xspace}
\newcommand{\etal}{et al.\xspace}

\title{Reflections on Traceability for Visualization Research}

\author[J. Rogers, D. Akbaba, J.\,S. Brown, A. Lex, \& M. Meyer]
{\parbox{\textwidth}{\centering 
J. Rogers$^{1}$\orcid{0000-0002-9568-469X}
D. Akbaba$^{2}$\orcid{0000-0001-9419-3402}
J.\,S. Brown$^{3}$\orcid{0000-0001-5642-8346}
A. Lex$^{4,5}$\orcid{0000-0001-6930-5468}
and M. Meyer$^{6}$\orcid{0000-0002-8971-6245}}\\
{\parbox{\textwidth}{\centering 
$^1$Idaho National Lab, USA\\
$^2$KTH Royal Institute of Technology, Sweden\\
$^3$University of Edinburgh, UK\\
$^4$Graz University of Technology, Austria\\
$^5$University of Utah, USA\\
$^6$Linköping University, Sweden}}
}

\begin{document}


\maketitle

\begin{abstract}
Decades of advocacy for reproducibility and replication have advanced open, transparent practices in the sciences. However, traditional notions of reproducibility fit poorly with design-oriented visualization research, where insights emerge through subjective, situated, and iterative work.  So how can we ensure rigor and transparency in processes that are inherently unreproducible? To introduce transparency in design-oriented research, we propose to focus on traceability: surfacing the origin and development of research contributions based on rich sets of artifacts documenting the design process. We investigated traceability through a collaborative autoethnographic reflection that builds on several years of work exploring ways to make design-oriented research transparent. This exploration includes an experiment to build a tool to support traceability, which we called tRRRacer. The tRRRacer tool provided a testbed for us to operationalize the three tenets of a traceable process: (1) Record abundant, annotated artifacts representative of research activities; (2) Report curated research threads that articulate rationale and evolution of the process, allowing others to (3) Read via interfaces that help retrace claims and assess plausibility. Reflecting on our experiences, we contribute a theorization of traceability and reflections on how we might support it.


\end{abstract}

\section{Introduction}

Reproducibility and replication of empirical scientific research are prominent focuses and concerns in computer science~\cite{national2019reproducibility}. Reproducing and replicating results provides a research community with the means to validate and trust their truthfulness. The visualization research community has championed reproducibility and replication for decades, producing a shift towards open research practices, more access to data, preregistration of studies, and an increased valuing of replication  studies~\cite{fekete2020exploring,besancon2021publishing,nosek2018preregistration}.

However, the field of visualization --- like computer science more broadly --- is increasingly methodologically diverse with a plurality of epistemic foundations and approaches. Today, many visualization methods of inquiry include subjective and reflective practices, such as design methods and qualitative analysis, making them ill-suited to traditional reproducibility and replication requirements. In considering subjective and reflective research, scholars in the social sciences and design research communities instead advocate for transparency and scrutinizability so that others can assess the credibility, plausibility, and rigor of the research results~\cite{lincoln1985establishing,tracy2010qualitative,zimmerman_role_2008}. But what does it mean for design-oriented visualization research to be transparent? How do we record and report reflective processes? What is --- or could be --- the evidence of iterative, subjective, and reflective research methods? And how do we make this evidence legible to others?

In this work, we propose a theoretical framing called \textit{traceability} to support the transparency and scrutinizability of non-reproducible visualization research. This framing is grounded in the concept of material traces from science and technology studies (STS), which theorizes that objects in the world are embedded with traces of their making. From this position, we articulate a set of concepts for visualization research --- research artifacts and research threads --- for making design and reflective work in research legible to others. 
This theoretical framing originates from a collaborative autoethnographic reflection~\cite{lily2025autoethnography} on a design experiment in which we created a tool for supporting traceability. Called tRRRacer, this tool was designed to support the core tasks of tracing: recording, reporting, and reading. Although our experiment did not result in a tool for public dissemination, it did provide us with new ways to think of how we might support traceability in the future.

The core contribution of this work is a theoretical conceptualization of traceability intended as a generative framing for reasoning about recording and reporting visualization research.
We ground this conceptualization in theories from STS as well as  
a design experiment in which we designed, developed, and tested a new tool for recording and reporting on visualization research. Reflecting on our successes and failures in the experiment, we propose speculative future directions to support the traceability of visualization research and question how much transparency is enough. We see this paper as part of a continuing discussion on visualization design, open practices, and rigor across an increasingly diverse set of visualization research methods.

\section{Background}

The ability to reproduce and replicate scientific results is a hallmark of good science, providing a means to validate the truthfulness of studies and, ultimately, to establish confidence and trust to build upon others' prior work. The inability to reproduce and replicate vast numbers of published scientific studies, however, has gained attention in both the scientific literature and popular media. In response to this \textit{replication crisis}, the US government commissioned a study in 2017 to assess the reproducibility and replicability of scientific and engineering research. 
The study's final report defines reproducibility as the ability of a second researcher to recompute original results through the ``availability of data, code, and methods that make that recomputation possible''~\cite{national2019reproducibility}. Replication is defined similarly, but using independently obtained data. 
The report acknowledges, however, that not all scientific research is replicable, recognizing that ``reproducibility and replicability are not, in and of themselves, the end goals of science, nor are they the only way in which scientists gain confidence in new discoveries''~\cite{national2019reproducibility}.

A study by Raghupathi \etal looked at reproducibility characteristics across a corpus of computer science studies and found that the multidisciplinary nature of the field, along with the development and validation of artifacts in socio-technical contexts, contributes to problems with reproducibility
~\cite{raghupathi2022reproducibility}. A critical issue their study highlights is the subjective and largely undocumented nature of technology design and development: practices that are poorly understood and do not fit neatly into normative frameworks for verifying the validity of hypothesis-driven scientific research.

Visualization researchers have a long history of advocating for making visualization research reproducible~\cite{silva2007provenance,fekete2020exploring,besancon2021publishing}, as well as in developing provenance tools for supporting the reproducibility of visual data analysis~\cite{xu2020survey}. 
\rev{Provenance tools focuses on addressing the questions \textit{what happened} and  \textit{when} in an interactive visual analysis session~\cite{ragan2015characterizing,cutler2020trrack,gratzl2016visual,silva2007provenance,stitz2018knowledgepearls,chang2009defining}. This work is important in the context of this paper for understanding current practices and tools for tracing the iterative work of data analysts and creating readable traces for others~\cite{gratzl2016visual,wang2022forseti}. 
This work, however, focuses on making visual analysis processes transparent during the \textit{use} of a visualization tool, while the work we present in this paper highlights the unique considerations and challenges of tracing the subjective and undocumented practices of visualization \textit{design}.  }

Visualization researchers have argued that some approaches to 
research --- such as visualization design studies~\cite{sedlmair2012design} that focus on creating visualization tools for real-world contexts --- are not only inherently unreproducible, but attempting to make them so would undermine the rigor of the design-oriented research itself. Alternative requirements focus instead on making research \textit{transparent} and \textit{scrutinizable} such that others can make judgments about the appropriateness of methods, quality of evidence, and reasonableness of conclusions~\cite{meyer2019criteria}. Other visualization researchers have proposed pragmatic approaches to enabling transparency of some design 
processes, such as \textit{literate visualization} that calls for explicitly including rationale for decisions made while iteratively coding visualization tools~\cite{wood2018design}. 

The arguments and theories behind calls for transparency in visualization research 
--- as well as in HCI more broadly~\cite{wacharamanotham2020transparency,niksirat2023changes} --- 
build from long-established research norms in the social sciences, and for qualitative research in particular~\cite{lincoln1985establishing,tracy2010qualitative}. 
Software systems for coding qualitative data help enable transparent reporting of analysis~\cite{lu2008rigor}, and reflective memoing during the research process makes researchers' subjectivity and biases more visible~\cite{birks_memoing_2008,charmaz2006constructing}. Additionally, \textit{audit trails} are a widely used method for supporting the evaluation of qualitative studies by giving auditors access to a researcher's methodology, decisions, and analytical processes to confirm a study's findings~\cite{carcary2009research}. We observe that qualitative researchers have already established rigorous methods to trace their non-reproducible findings, unlike visualization design researchers, who are the focus of our efforts. An exception is the work by Rogers et al.\ which experimented in building audit trails for visualization research by providing access to study materials via web-based tools~\cite{rogers2020insights}. 
We found existing guidance on building audit trails to report on interviews and observations, but not for transparently reporting on the design and development of technology artifacts.

Instead, work within the design research community has focused on acknowledging and communicating acquired design knowledge embedded within artifacts~\cite{zimmerman_role_2008}. As carriers of knowledge, designed artifacts are imbued with what a designer comes to know about needs in the world and ways of shaping materials into a desired object~\cite{cross1999design}. This knowledge is, however, opaque and requires explicit documentation about the underlying design rationale and decisions. \textit{Annotated portfolios} are an approach for revealing design insights from the abstraction of artifacts, designed for specific situations, into generalized knowledge~\cite{gaverbill_annotated_2012}. This is often accomplished by bringing together a collection of artifacts and making explicit the designer's particular knowledge embedded in each of them, and how that knowledge generalizes across the collection. While annotated portfolios provide insight into the importance, novelty, and innovativeness of artifacts, they do not reveal the underlying processes of how these artifacts came to be.

In summary, visualization researchers have acknowledged and embraced requirements for making their work reproducible, yet not all visualization research is inherently so. In particular, the design of visualization interfaces and tools requires some amount of subjectivity and reflection, making these design practices ill-suited to reproducibility and replication requirements. More recently, researchers have instead proposed that these studies be \textit{transparent} and \textit{scrutinizable}, building arguments from similar calls in the fields of social science and design research. These fields, however, do not account for the specific challenges, needs, and processes of designing visualization tools, leaving a gap in knowledge about \textit{how} to make visualization research transparent and scrutinizable for others. Our work, described in this paper, is an exploration of how to fill this gap.

\section{Traceability}

The goal of our work is to make visualization research \textbf{traceable}: that \textit{the doing and learning from research practices are transparent, scrutinizable, and legible}. 
We consider a traceable research process to be one that supports others in understanding what was done and why, and how the findings, insights, contributions, and conclusions of the research came to be. 
Importantly, tracing a research study is something that is done by \textbf{the reader} of a research report as they seek to understand what was done and why, with support from \textbf{the researcher} through the evidence and descriptions made available.

\rev{
Our conceptualization of traceability draws from scholarship on material traces from STS -- the idea that 
objects are imbued with cultural and social knowledge that can be read through their material properties. Material traces capture this link between a phenomenon and the marks it leaves, for example, footprints in sand left from a walk on the beach ~\cite{offenhuber2019data}. More broadly, these marks can support scrutiny of what happened and why~\cite{offenhuber2019data,dourish2022stuff}.
History of science scholars describe science as a process of trace-making, where traces serve as the objective evidence left behind~\cite{daston2021objectivity}. More recent theorizing emphasizes that traces are not simply found but become meaningful through interpretation when someone engages with those marks~\cite{barad07meeting}.
}

\rev{Building from these foundational concepts, we treat traceability as an intentional research practice, in which researchers must deliberately record and richly describe evidence so that readers can reconstruct what was done and why. 
This perspective aligns with transferability in qualitative research, where readers assess how insights might carry to their own context, supported by \emph{thick description} --abundant, context-rich accounts of the research setting, methods, observations, and reflections~\cite{geertz2008thick,lincoln1985establishing,meyer2019criteria}. }
\rev{Here, thick descriptions can take the form of contextual and interpretive notes akin to qualitative fieldnotes and analytic memos that capture what was happening, why an artifact matters, and how interpretations and decisions evolved, leaving an auditable trail beyond the artifact alone \cite{birks_memoing_2008}.
}

We consider traces as something interpreted by readers as they engage with marks left behind from the research process --- the evidence --- that are made available by the researcher. This perspective has two important implications. First, like critiques of big data that argue data is meaningless without context~\cite{boyd2012critical,gebru2021datasheets}, the evidence collected from a research process requires context, description, and annotation to be meaningfully interpreted by others. 
Second, a reader's sense-making 
of a trace is a type of knowledge- and meaning-making that is situated within a reader's own knowledge, experiences, biases, and history~\cite{haraway2013situated,barad07meeting,dignazio2020data,akbaba2024entanglements}. Taken together, these two implications mean that a researcher cannot control \textit{exactly} how a reader will interpret the evidence they make available, but they \textit{can} make their research process more legible through abundant evidence that is thickly annotated. Thus, for researchers, the goal of traceability is to leave abundant marks that are rich, vibrant, and authentic, such that the reader’s interpretation of the traces reflects the nuance of the underlying processes as closely as possible.

But what, more specifically, is important to trace from visualization research? We propose two concepts --- research artifacts and research threads --- to address this question, as well as three design requirements of recording, reporting, and reading to support tracing in practice. 

\subsection{Research Artifacts}

A visualization design process consists of a set of research activities.
When designing visualizations, these activities can span a myriad of possibilities, from conducting
an interview, collecting measurements during a user study, or iterating on a visual encoding, to ideating on a whiteboard, coding a feature in a software prototype, or running an analysis script. Supporting the traceability of these activities gives readers insight into \textit{what} was done during the study.

To make activities traceable, we propose a specific conceptualization of \textbf{research artifacts}. 
Research artifacts are \textit{the marks left behind from a design process; they are the evidence of what was done.} Different types of artifacts are evidence of different visualization design activities: visualization tools reflect how a researcher interpreted a data analysis need in the world; transcripts from interviews capture how a researcher engaged with a domain and the people in it; images taken during a participatory workshop show the activities that a researcher used to engage participants, and the resulting ideas that emerged; analysis code and scripts capture the statistical models a researcher used to interpret data; and much more. It is from an abundant and rich collection of artifacts that a reader can interact and interpret the trace of the activities a researcher engaged with to better understand the research process. 

Constructing rich, vibrant, and authentic traces of these activities requires more than just a dump of numbers or a pile of screen-grabs. Although these artifacts are imbued with the details of the research activities that produced them, these details may not be legible to anyone other than the researcher who recorded them~\cite{cross1999design,gaverbill_annotated_2012}. 
Annotating research artifacts with rich details about the context of the activity can provide a more vivid picture of the processes that led to the artifacts~\cite{bowers_logic_2012}, literate visualization captures rationale during visualization development~\cite{wood2018design}, and computational notebooks make data analysis legible~\cite{rule_exploration_2018}.

\subsection{Research Threads}
The insights that a researcher gains through designing a visualization come from exploring, testing, and learning across \textit{multiple} activities. 
For example, a researcher interviews collaborators about their domain, learning about where their data comes from and the kinds of analysis they hope to accomplish. The researcher then sketches ideas of possible visualizations, going back to the interview to motivate and justify different design ideas. Taking the sketches back to the collaborators for feedback, the researcher collects their thoughts on the feasibility and usefulness of the ideas. These three activities --- an initial interview, an ideation session, and a feedback interview --- are not isolated but are meaningfully connected by their involvement in a researcher's process of producing a new visualization technique.
Supporting the traceability of this learning process is important for communicating \textit{how} the researcher's insights and conclusions came to be. 

Traceable visualization research requires making research activities and their connection to research results transparent and scrutinizable.  
We propose \textbf{research threads} as a mechanism for supporting readers in tracing how a researcher came to conclusions, designed a new tool, or gained some insight over the course of a study. More specifically, a thread is a \textit{curated collection of artifacts that are evidence of a researcher's learning process that trace a research result from its emergence.} Like artifacts for constructing traces of activities, threads for constructing traces of learning require annotation and descriptions to make the meaning of the relationships legible to readers: to explain why artifacts matter and how they fit together. 

\subsection{The Three Rs: Design Considerations for Traceability}

Supporting traceability \rev{requires the consideration of three critical tasks} --- \textbf{recording}, \textbf{reporting}, and \textbf{reading} --- that mediate the communication of research activities between researchers and readers. The first two of these tasks --- recording and reporting --- focus on supporting researchers as they document their processes, while the last --- reading --- supports consumers of a research paper (reviewers or readers) in the understanding of traces, i.e., how an insight or result emerged from the process. Crucially, these 3Rs support the transparency and scrutinizability of visualization research.  

\noindent\textbf{R1: Recording} Traceability fundamentally builds from an abundant and diverse collection of artifacts, which researchers must thoughtfully produce, collect, store, and annotate. Recording of research activities produces artifacts, which can be considered a making activity that creates persistent evidence of an otherwise ephemeral process. Tools and processes that support recording need to accommodate a wide range of research artifacts, as well as provide mechanisms to annotate and curate them. 

\noindent\textbf{R2: Reporting} Researchers can construct threads from a collection of artifacts through a process of \textit{reflection-on-action}~\cite{schon2017reflective}. These threads report on the researcher's actions and reasoning during the research process to facilitate the validation and auditing of findings by others. Reporting requires a researcher to access and review artifacts, to thread together a curated set of artifacts, and to describe why and how these artifacts fit together.

\noindent\textbf{R3: Reading} 
Finally, research activities and threads must be accessible, legible, and scrutinizable by external readers, such as students, colleagues, reviewers, collaborators, and others.
Readers should be able to engage with threads and artifacts to understand and validate the research process, as well as assess the plausibility of the derived outcomes. Traceability rests on readers' understanding of how research findings came to be through access to threads and artifacts.

\section{The tRRRacer Experiment}

\rev{In this section, we outline the design experiment we conducted to refine our understanding of tracing visualization research. We conducted our experiment as a collaborative autoethnographic study, grounded in years of previous work exploring transparency practices within design study research. This study, like other recent collaborative autoethnographies in visualization~\cite{lily2025autoethnography} and HCI research~\cite{ciolfi2025doing}, provided us the opportunity to use our first-hand experiences as a starting point for new knowledges and understandings~\cite{howell2021cracks}.}

\rev{We include a description of our experimental journey, details of the tool we developed called tRRRacer, and what we learned through it all. We supplement this description with deep links to artifacts [\parfig{artifact}], activities [\parfig{activity}], and threads [\parfig{thread}] to enhance tracing of our journey by you, dear reader. }

\subsection{Process}

\rev{The origins of the design experiment were rooted in our collective struggles in reporting on design-oriented, reflective visualization research. We initially explored ways to report on design studies using audit trail-like supplemental materials where we reported on research activities in a timeline view [\externallink{https://vdl.sci.utah.edu/IEFramework/suppmats.html}, \externallink{https://vdl.sci.utah.edu/CVOWorkshops/audit/}].
In these projects, we did not focus on recording a diverse collection of artifacts; instead, we predominantly documented them through meeting notes and screenshots. In a later design study, we focused on newly proposed rigor criteria~\cite{meyer2019criteria} and worked to collect an abundance of evidence [\externallink{https://vdl.sci.utah.edu/trrrace/}]. Here, our artifact collection included emails, annotated literature, sketches, photographs, transcripts, and more. Our reporting included deep links to individual artifacts directly in the paper~\cite{rogers2020insights} as well as a timeline view of the full collection. }

\rev{These early experiments focused on how to visualize a collection of research artifacts, relying on timeline-based views. Each project increased our awareness of the need to systematically and consistently record research artifacts, as well as to consider the reading experience. Building on these early efforts, we wanted to extend these ideas toward practical and technical support for traceability. To better understand the needs, challenges, and opportunities for supporting tracing, we embarked on a design experiment to build and test a new tool called tRRRacer. 
We conceived tRRRacer as a tool to support the 3Rs of tracing: recording, reporting, and reading.}

\rev{Early in the experiment, we identified several internal research projects that could serve as a testbed for both gathering ideas and eliciting feedback. 
The first was a retrospective look at the data from our published design study in which we had digitized an abundant collection of artifacts~\cite{rogers2020insights}.
The second project was the development of tRRRacer itself, where we used tRRRacer as it was being developed to document our design process and experimentation. 
The third project was an ongoing interview study about visualization design study collaborations~\cite{akbaba2023troubling}, and the fourth was an ongoing design study that remains unpublished. 
Through these four projects, we sought to identify commonalities between the recording and reporting needs of different researchers in different types of visualization research projects. We used the retrospective project as a starting point in the development of tRRRacer, and incorporated the other ongoing projects once the tool had sufficient features for use.}

\rev{The tRRRacer experiment took place from the spring of 2021 until the fall of 2022. During this period, the research team met every week, discussing the status of the tool, receiving feedback on its use, and brainstorming for future directions. Rogers and Brown led the development of tRRRacer, with Akbaba serving as an additional user of the tool.
This period of active development and use lasted until the main researcher on the project, Rogers, completed her dissertation [\externallink{https://vdl.sci.utah.edu/publications/2023_thesis_rogers/}], and as a team, we drafted an initial manuscript of our findings. Our initial manuscript was submitted but not accepted for publication [\externallink{https://trrracer.netlify.app/?view=paper}]. At this point, the team had moved on to other research projects; the lead author started a new position; and the tracing research was paused.}

\rev{This pause provided us time and space to rethink our experiences with tRRRacer and our transparency needs across a variety of research projects. During this period, Meyer gave a number of invited talks that included discussions of tRRRacer and our initial theorizing about traceability. Positive feedback from these discussions encouraged us not to abandon the project, and in the spring of 2025, we began internal discussions to rethink our theoretical framing and recommendations for future tracing research. Over the course of several months, Rogers, Lex, and Meyer met to discuss the tRRRacer experiment and to reflect on what we would do differently next time. Along with Akbaba, we developed a digital whiteboard of reflections [\trrracer{jen}{overview}{activity}{47a5f682-70e3-478c-a635-f3b68f8619990}] that we grouped into a final set of lessons learned and future considerations we present in this paper.  }

\subsection{The tRRRacer Tool}

\begin{figure*}
\centering 
    \includegraphics[width=\textwidth]{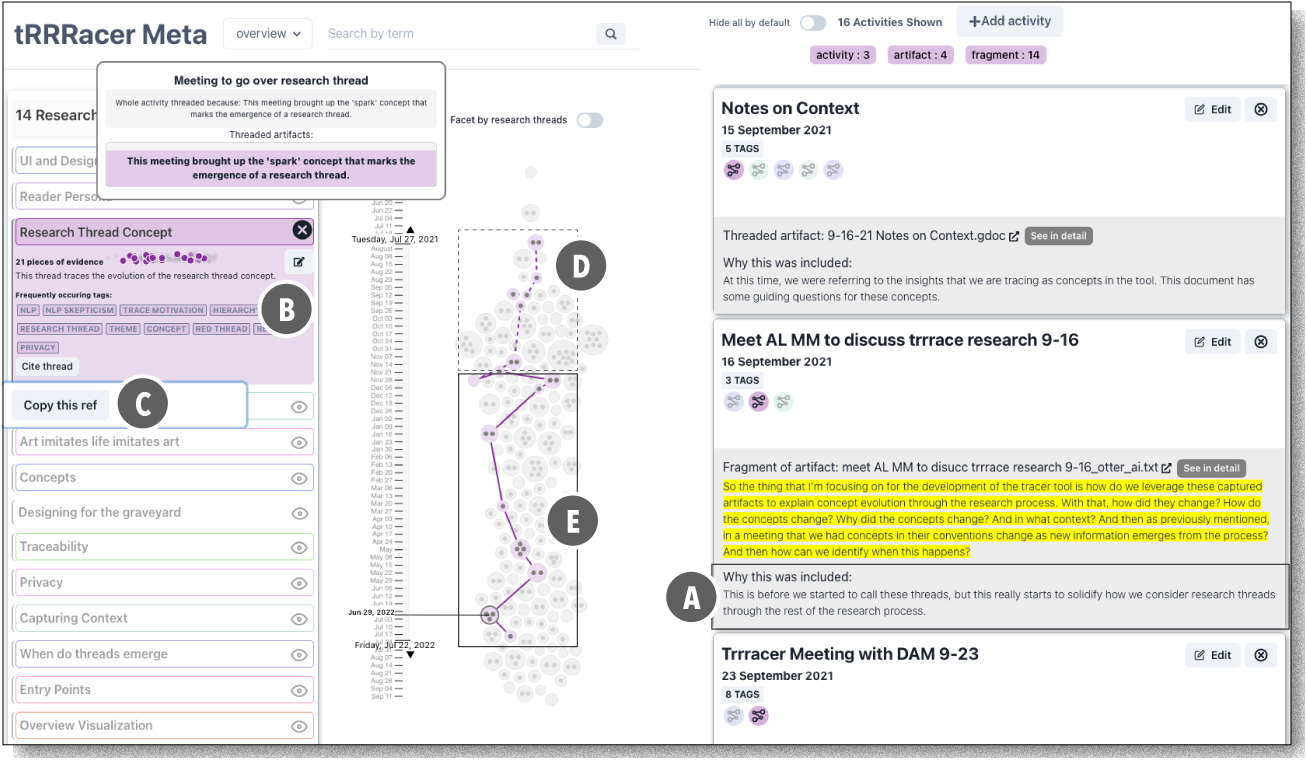}
    \caption{The tRRRacer desktop interface. (A) The detail view shows artifacts collected in the selected thread. When adding a piece of evidence to a thread, the researcher is encouraged to also provide a rationale for why it contributes to a given thread.  
     (B) A list of all threads is shown in the sidebar, which can also be used to create new threads. 
     (C) Researchers can copy an automatically generated thread citation to use in a paper. The citation looks like this: \texttt{\string\trrracer\{project name\}\{type of view\}\{granularity of evidence\}\{id\}}
     (D-E) Threaded activities are shown as connected in the overview bubble visualization. (D) The dotted lines linking activity bubbles represent activities threaded retroactively; (E) solid lines indicate that activities were added after the thread was created.  }
    \label{fig:thread-tool-view}
\end{figure*}

\rev{
We built two versions of tRRRacer: a desktop version built with Electron for use by researchers during their work, and a read-only website 
[\trrracer{jen}{overview}{thread}{ba27001f-5d28-44eb-b507-bd7b0ff0be7a}] for readers to trace a research process. Both applications share a single codebase and connect to the same Google Drive repository of notes and artifacts; the Electron app exposes the full authoring interface, while the Reader web app reuses the same logic but removes all editing functionality so readers can only browse.}
You can try tRRRacer Reader here: \url{https://trrracer.netlify.app/}

We designed tRRRacer as an overview + details-on-demand interface, shown in Fig.~\ref{fig:thread-tool-view} and Fig.~\ref{teaser}.
Activities (light bubbles) and their nested artifacts (dark bubbles) are laid out over time in the middle of the interface to reveal bursts, lulls, and clusters of research work. This overview view is flanked on the left by a view detailing the available threads. When a thread is selected, it is shown in the middle overview as a connected path through a subset of activities. In the desktop tRRRacer (Fig.~\ref{fig:thread-tool-view}), the right side view shows artifacts in detail in the context of
a selected activity. In tRRRacer Reader (Fig.~\ref{teaser}), artifact details are shown in a pop-up window, with the right side view showing a research report that contains embedded, deep links to artifacts, activities, and threads. All views are linked via user interactions and selections. In the following subsections, we describe our specific design decisions for supporting the 3Rs of tracing.

\subsubsection{Record an Abundant Collection of Artifacts}
Traceability relies on an abundant and structured collection of artifacts. To make captured evidence legible, tRRRacer requires the user to specify a type for every artifact recorded --- such as \textit{transcript}[\externallink{https://trrracer.netlify.app/.netlify/functions/download-gdrive-file/?folderName=jen\&fileName=Metting\%20with\%20JSB\%20for\%20tRRRace\_otter\_ai.txt\&raw=1}], 
\textit{sketchbook page}[\externallink{https://trrracer.netlify.app/.netlify/functions/download-gdrive-file/?folderName=evobio\&fileName=513.png\&raw=1}], \textit{memo}[\externallink{https://docs.google.com/document/d/1WgnCVOy7pvm\_g16jI7nqEa0M53nx1FbjkTml8PrCNaI/view?tab=t.0}] --- and a brief description. Artifact types add semantic context and enable filtering; annotations record why the artifact matters and what decision or insight it supports. In tRRRacer, artifacts --- notes, screenshots, recordings --- are attached to a design activity --- meetings [\trrracer{jen}{overview}{activity}{03db5e56-5181-4225-a0d7-87d9ebd24005}], workshops [\trrracer{jen}{overview}{activity}{684f69e7-327e-4992-bcf2-9525496cf9d1}], design sessions [\trrracer{jen}{overview}{activity}{9309dc93-c795-4be6-b2b7-98bad5230350}], interviews [\trrracer{jen}{overview}{activity}{f0b03554-09b6-43b3-8b28-dcb096e71c86}]. We integrated the Google Drive API to support the creation of files directly within the tool.

Activities and artifacts can also be tagged. Tags are a lightweight method for researchers to synthesize important information during recording, to leave breadcrumbs for later reflection, and to seed threads by linking related evidence across activities. Tags also support readers by enabling filtering and discovery during reading~[\trrracer{jen}{overview}{activity}{3d5b48cc-35b1-499d-b3dd-1d53471d7b84}, \trrracer{jen}{overview}{activity}{a83e9821-9327-448d-9843-680748ec56d5}].

\subsubsection{Report with Threads}
The activity of reporting creates a bridge from raw records to legible evidence. We designed tRRRacer’s reporting workflow to: curate artifacts into useful narratives, expose the rationale behind design decisions, preserve temporal and contextual links among activities, and prepare evidence so others can later retrace and scrutinize claims. All of these activities are captured in research threads. 

We designed research threads to curate and annotate evidence that traces how a research idea emerges and evolves [\trrracer{jen}{overview}{thread}{b973c840-f26c-4a21-99b3-b93e411d659c}]. Researchers create a thread when a concept becomes salient as an important insight of the project. 
To emphasize which artifacts predate the explicit recognition of an insight, we visually distinguish post-hoc from later additions to the thread (dotted vs.\ solid links; Fig.~\ref{fig:thread-tool-view}D--E). We ask users to provide a brief rationale whenever evidence is added to a thread (Fig.~\ref{fig:thread-tool-view}A) to encourage articulation of why the evidence matters at the moment of curation and to leave an explicit record for transparency. 
Researchers can highlight areas of text to include in a research thread, which we refer to as artifact fragments (as seen in this example thread [\trrracer{jen}{overview}{thread}{a158e0bb-ff85-42e4-add2-94de1087db92}]). tRRRacer also supports editable and mergeable threads. Editing prevents premature lock-in; merging supports abstraction by combining related threads into higher-level ideas.

\begin{figure*}
    \includegraphics[width=\textwidth]{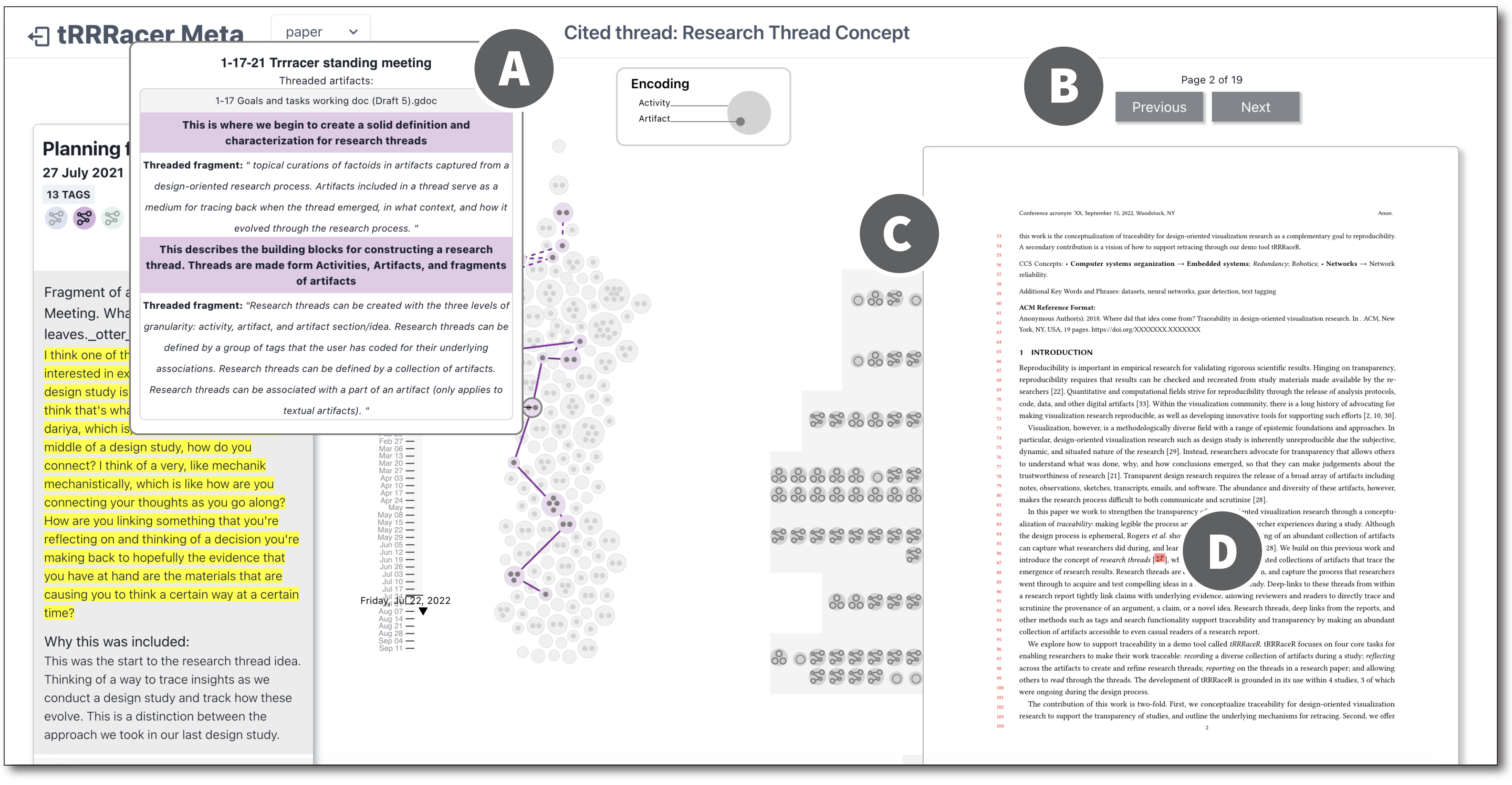}
    \caption{The read-only   
    tRRRacer Reader interface with a paper view. (A) The panel on the left shows relevant activities and artifacts for a selected citation. This example shows a tooltip explaining the rationale for artifacts' inclusion in the research thread for the `Research Thread Concept'. (B) A reader can navigate by page, reading the manuscript in the view, or (C) they can explore with the side annotations for links on the side. (D) An example of an inline deep link. Clicking on these navigates to the detail view for a given activity or artifact. }
    \label{teaser}
\end{figure*}

\subsubsection{Read and Retrace}

To support readers we built an explicit reading view -- tRRRacer Reader -- to support understanding of what happened and when, and to trace salient ideas via research threads. Researchers can copy an auto-generated citation for any piece of evidence (Fig.~\ref{fig:thread-tool-view}C) and embed it in a report. 
tRRRacer Reader then embeds a report view (Fig.~\ref{teaser}) that interacts with the rest of tRRRacer through the deep links specified in the report. tRRRacer Reader shows per-page deep-link previews (Fig.~\ref{teaser}C) so readers can scan where evidence is concentrated and jump directly to cited artifacts, activities, and threads. Clicking a tRRRacer link in an external PDF viewer opens tRRRacer Reader [\trrracer{jen}{paper}{artifact}{1-ynzmohY1iqCgi0Od5bA5bEUFtbJ-Jqr}].

tRRRacer Reader is read-only to protect evidence integrity and enable safe sharing beyond the authoring team. We set privacy defaults that hide activities marked as private and replace names with initials to reduce re-identification risk and align with requirements for double-blind reviews and potential ethical expectations (e.g., hiding full-text interviews) [\trrracer{jen}{paper}{thread}{523624cf-2fbd-4c00-ac69-c36b8a0f49bb}]. To support discovery beyond citations, we provide a time-based overview of activities with nested artifacts, answering \textit{“what happened, when?”} at a glance. Navigation is thread-centered: the left sidebar lists research threads, and selecting one highlights it in the overview to foreground the curated rationale connecting evidence to claims. Readers can move fluidly between paper-linked entry points and process-wide exploration to retrace claims at both narrative and procedural levels.

\subsection{Threads as Windows into the Research Process}
\rev{To make the concept of research threads more concrete, we briefly explore their affordances in the context of three vignettes. 
These speculative vignettes, derived from our internal testbed projects, are meant to explore how threads might support reporting on (a) the merging and branching of ideas, (b) dead ends and abandoned directions, and (c) the gradual surfacing of concepts over time. We offer these as an experiment for you, the reader, to reflect on both the opportunities and the challenges of tracing visualization research.
}

\subsubsection{Tracing Concept Development: Merging Threads}
To inform the initial design of tRRRacer, we constructed a retrospective research thread for our prior design study in evolutionary biology~\cite{rogers2020insights}. In that project, the evolutionary concept of \textit{convergence} emerged in initial meetings and sketches with our biology collaborators as a tentative, standalone direction for our tool design. As the project progressed and our understanding of the domain and stakeholders solidified, \textit{convergence} was gradually absorbed into a broader, more expansive concept we called \textit{patterns of evolution}, which we ultimately used as a design goal and reported as a central focus of our design study. The associated thread [\trrracer{evobio}{overview}{thread}{f991e941-1c53-4174-a47c-104ddd55fbdd}] connects early notes, sketches, and design reflections centered on \textit{convergence} to later design rationales and manuscript text, focused on \textit{patterns of evolution}.

This example motivated our support in tRRRacer for \emph{branching} and \emph{merging} operations for threads. Branching captures how lines of inquiry naturally split into parallel probes and alternative framings, while merging captures how recurring themes later coalesce intohigher-levell concepts. We speculate that making these research paths explicit can document idea growth, reduce redundancy, and preserve provenance.

\subsubsection{Tracing Abandoned Directions: Dead-End Threads}
Threads can capture abandoned research directions. An example of this is our thread \textit{NLP for Entry points} for this work, constructed early in the research process when we attempted to use off-the-shelf NLP methods to surface promising starting points in a large corpus of project notes. Despite initial enthusiasm, we ultimately concluded that the recommended leads did not meaningfully guide our work and decided to abandon this direction. The associated thread connects initial brainstorming notes \textit{``could we auto-suggest leads?''}, prototype screenshots, pilot evaluation notes, and later reflections where we articulated why the approach was not worth pursuing further [\trrracer{jen}{overview}{thread}{5c1a4d2a-7886-4c67-b0e8-fe005d1a6c95}].

Documenting failed avenues in threads has the potential to counter hindsight and survivorship
bias, showing that final claims arose from a systematic inquiry rather than cherry-picking. Dead-ends can expose alternatives considered and the rationale for abandoning them, which can strengthen the credibility of the chosen approach. They also improve transferability: readers can judge boundary conditions and avoid repeating unpromising paths in different contexts. Finally, unsuccessful artifacts could become seeds for later ideas, and we speculate that recording them in a thread could preserve reusable knowledge and starting points for future explorations.

\subsubsection{Tracing a Core Concept: Evolving Threads}
Recording and visualizing threads can surface when and how ideas evolve over time. 
To make our process for this paper traceable, we created threads documenting the development of central concepts that we grappled with as a research team, including the concept of research threads [
\trrracer{jen}{overview}{thread}{b973c840-f26c-4a21-99b3-b93e411d659c}].
This thread illustrates our evolution of thought around research threads: from its first mention as an idea for creating \textit{``traces of a process as we go''}[
\trrracer{jen}{overview}{activity}{6fa07668-43d6-4d9b-ae3e-c74c9c2ee86f}], to the inclusion of tags as a helpful tool for \textit{``pre-threading''} [
\trrracer{jen}{overview}{activity}{f33cf2a3-48dc-4f49-8aa3-3d7e12e384a1}], and what information should be included within a thread. 

Another thread reveals that we were plagued with the recurring question: \textit{``when should you begin threading?''} [
\trrracer{jen}{overview}{thread}{e1d0fa29-b77b-479a-bb2f-421b974e9d84}]. 
This lingering question marked an interesting tension for us throughout the design experiment. We initially theorized that threads could support reflection-in-action, and yet we struggled to create threads \textit{during} the research process. It was only during writing that we could more easily generate threads. We return to this tension in the following section.

\subsection{Lessons Learned in Traceability}
\label{sub:lessons}

Developing the tRRRacer tool was a design experiment that gave us the opportunity to explore the challenges and opportunities of supporting tracing of visualization research. 
Through cycles of reflection, we distilled our learning about traceability into a set of lessons, drawn from both our experiences using the tool and the insights it enabled about our own research practices. We detail those lessons here. 

We note that we would not build tRRRacer again and do not plan on maintaining or expanding it. However, our experience with developing a singular tool for recording and reporting was insightful both in terms of how to support traceability, as well as in how we might build research tools that last.

\subsubsection{Successes}
One clear success of the tRRRacer experiment was our attention to artifact curation. Across the four experimental projects, we focused first and foremost on recording abundant artifacts --- sketches, meeting notes, emails, screen shots, reflections, and more. We began to approach all of our research activities with an eye towards documentation, and we made time to systematically and regularly reflect on why those activities mattered. 
The tRRRacer tool provided curatorial features beyond a storage space for digital files; instead, it oriented our recording processes toward organization, annotation, and externalizing how artifacts related to other concepts within the research project. 
Both student authors on the paper, Rogers and Akbaba, took these recording practices into their later research projects, where they used commercial tools to continue recording abundant collections of research artifacts. These practices of attending to context through tags and annotations of abundant artifacts have continued to support a nuanced and rich view of how research ideas are entangled and develop over time, serving as important reminders where our memory fails.

As part of artifact curation, using tagging was, and remains, important for discoverability. We frequently used the tags in tRRRacer for reviewing artifacts as well as for creating threads, and are continuing the practice of tagging artifacts in our current use of commercial tools. Additionally, the tags enabled a quick form of reflection-in-action: they are a best guess in the moment of why an artifact matters. While we struggled with creating threads in tRRRacer in the middle of projects --- our more intentional feature for reflection-in-action --- tags were lightweight and familiar. We discuss our difficulties creating threads in Section~\ref{sec:challenges}.

Beyond simple tags, tRRRacer also enabled reflection-in-action through explicit foregrounding of annotation practices. In particular, annotating research artifacts instilled a process of articulating why an artifact, and thus activity, was important within the research process. Annotating artifacts became akin to reflective journaling. 

We found that good annotation of artifacts and threads was critical to making threads readable. Our review of threads we created in the four projects required annotations to make sense of the artifacts by team members not involved in the projects. Annotations of threads also left us with a feeling of reading through a research journal. In short, a dump of artifacts is not enough. Evidence without context is insufficient for tracing a research process.

Our final point of success within the tRRRacer experiment was the use of deep links within a report. \rev{Inspired by positive feedback on our use of deep links in one of our grounding design studies~\cite{rogers2020insights}, we have continued to use this mechanism for reporting in more recent work~\cite{cutler2026revisit,akbaba2026thinking}. }In the tRRRacer experiment we imagined taking this even further, extending the idea to include links to curated threads. Placing access to research evidence directly at the point it matters in a report provides a new way for readers to engage with supplemental materials. It makes the scrutiny of materials easy for a reader and adds a layer of richness to otherwise static and flat PDF paper formats.

\subsubsection{Challenges}
\label{sec:challenges}

For all our successes with tRRRacer, there were also a number of challenges that we confronted. First and foremost were the technical ones. Our decision to build a tool that supported recording, reporting, and reading in a single interface led us down a path of significant engineering efforts that mirrored efforts being taken by companies of commercial documentation tools like Notion. Furthermore, our desire to support a broad range of research artifacts --- from JPEGs to Google Docs --- added additional engineering efforts in order to accommodate the various file formats and document APIs. And once we started using the tool across four projects, backward compatibility became a challenge for any significant changes we wanted to make to the tool. 
In the end, we found ourselves in a quagmire of software engineering needs and an increasingly complex tool that we would need to maintain. These technical challenges made it difficult for us to focus on the interactive and visual features of tRRRacer, and also difficult to change course when our feedback indicated the need for different approaches. 

We were unable to fully realize anonymization features due to both technical and conceptual challenges. Manual anonymization of artifacts is time-consuming and also results in artifacts that are more difficult to trace by the researcher. We wondered whether we should maintain two versions of the artifact collection, one anonymized and one not, and if so, how should we implement this? Or, should we try to codify an anonymization process in the tool that anonymizes automatically in reading mode? We found the desire to protect the privacy of our team and collaborators in tension with our desire to encourage rich and transparent research documentation. 

We also struggled technically and conceptually with how to maintain long-term persistence of research evidence. We were using tRRRacer to curate, manage, and store the artifacts and threads from several projects, but due to the increasing complexity of the tool, we could not imagine a long-term maintenance plan for keeping deep links in reports available in perpetuity. Here, too, we struggled to decide whether to focus on making the evidence scrutinizable through interactive visualizations in tRRRacer or persistent and accessible through a simple dump of data. 

Despite our awareness of the need to accommodate diverse research and documentation practices, our commitment to a single tool enforced the use of certain formats, tools, and concepts. For example, long text documents from meeting transcripts or interviews could contain multiple ideas involved with multiple threads. How to subset these text documents for threading in tRRRacer required specific documentation practices, which differed depending on whether the researcher documented in plain text files or Google Docs. Here again, we were stymied by the engineering required to realize our ideas. 

More conceptually, we found that creating threads as a project is developing --- a form of reflection-in-action --- was extremely difficult and rarely productive. We had imagined researchers creating threads as new ideas emerged, ideas that might merge together or branch apart, or even die altogether. We imagined a visualization that could show how learning happens within a project through a visualization of all the threads. In practice, however, we found threads difficult to create during projects; it turns out that knowing what will be most meaningful or what could develop into higher level ideas is incredibly difficult in the moment. Threads we did create during projects tended to be superficial and unstable upon later reflection. It was much easier to thread together artifacts and annotate their meaning in our retrospective project, where we were looking to tell the stories of our findings. 

As a final challenge, we found across all of the projects in our experiment that recording and reporting were incredibly time-intensive. Although curating and annotating research artifacts is a practice we are maintaining, we use these curated collections largely for internal use and not for transparent reporting. During our tRRRacer experiment, we experienced collecting artifacts and creating threads as an additional, lengthy step without clear benefits or demands from the research community.

\section{Discussion and Future Directions}
\label{sec:recommendations}

In this section, we put forward several ideas of how to move traceability forward in the visualization research community based on learning through the tRRRacer experiment. These ideas ask us to rethink the role of large, complex tools; to imagine threads as stories about our research; to embrace reflection throughout our research practices; and to reconsider whether more is always better.

\subsection{Process and Persistence}
\label{sec:pandp}

Our findings from the tRRRacer experiment, both the successes and challenges, point to the need for processes over tools for traceability. Instead of trying to build yet another tool to support documenting research projects, we envision a set of guidelines and standards for the community that detail what kinds of artifacts to collect for different kinds of activities, what metadata and annotations to contextualize the artifacts, and when and how best to record. This approach would leave open the opportunity for researchers to use the tools that are most suited to their needs while providing them with actionable goals for what and how to record. These guidelines and standards, however, will require further experiments by visualization researchers in a range of settings who are committed to exploring traceability of their own work. 

Additionally, we advocate for the use of simple, time-tested file formats for storing research artifacts. Formats like JPEG, MPEG, PDF, and plain text are all likely to be supported for many years to come. Restricting our documentation to these types of formats will ensure evidence is accessible to others using a range of tools without dependencies on singular companies. While interactive visual interfaces have been great for supporting the needs of many, we argue that maintenance and persistence of tools is no longer something we should ignore. Simple formats, along with guidelines and standards for recording practices, provide a common basis that will stand the test of time.

Preservation of evidence has long been a priority for advocates of open science~\cite{rieger_sustainability_2012}. With the rise of open data in the sciences, researchers began looking for robust ways to maintain data and findings, and repositories like arXiv and OSF came online. While we think the practices in the open science community of using online repositories are reasonable, current geopolitics and the restructuring of academic research give us pause. It is clear that open science repositories that rely on governmental funding, or any sort of funding, are vulnerable to change. As such, we advocate for redundancy when storing artifacts and threads. Researchers may decide to use established preprint repositories, while also including the collection of evidence with publishers and on their personally maintained websites, or commercial ones like GitHub. If using deep links to research collections in a paper, we could imagine a persistent version of the paper that could be downloaded along with the collection that has links to locally stored files, or that links to an attached appendix section. Nothing is forever, but we may be able to hedge our bets through redundancy.

\subsection{Threads as Stories}

Although we found thread creation to be difficult \textit{during} a research project, it was a useful mechanism for reflection-on-action in later stages of projects. The process of creating threads is an opportunity to revisit evidence about how important and salient ideas came to be. We recommend that threads be a focus of reporting and that they be created during the process of writing up research results. In this way, threads serve as an alternative view into a research project, complementary to more traditional text-based reports. 

Placing thread creation in the reporting phase of research opens up the idea of threading as storytelling. This perspective acknowledges threading as an act of curation rather than completeness. Threads-as-stories also emphasizes the interpretive, subjective, and selective nature of what aspects of the research process are made visible, and how. Thus, it can be argued that threads are biased reconstructions, but we would counter that \textit{all} reporting on research inherently has this quality. Positioning threads as stories acknowledges the subjectivity of reporting and gives space for thoughtful recording and reporting practices to reduce biases, such as the use of thick descriptions~\cite{geertz2008thick,birks_memoing_2008} and abundant evidence~\cite{meyer2019criteria}.

We see interesting future work in how we might design authoring tools for thread creation. Similar to work in provenance for crafting stories about insights from visual analysis sessions~\cite{gratzl_visual_2016}, these tools would provide researchers opportunities to explore through past research activities via a collection of artifacts, and to link them together in ways that reveal a thinking process. Designing these tools would be an opportunity to explore anonymization strategies, such as outputting a curated collection of artifacts associated with a thread that have been sufficiently anonymized. An interesting design question is how to best visualize threads and their associated contextualizing annotations and metadata, and how to do so in a way that is persistent. 

\subsection{Other Benefits of Traceability}

This paper emphasizes how traceability strengthens rigor in the research process.  
Yet when we first initiated this project, we also envisioned that the act of recording and reporting could actively improve research outcomes through reflection-in-action.  
We asked whether routine engagement with our prior notes, ideas, and decisions might spark new insights, reveal blind spots, or accelerate refinement.  
We likewise anticipated that maintaining rich artifacts and traces would simplify reporting --- making paper writing easier because arguments and ideas would already exist in a well-documented form.  

In practice, achieving this reflection proved challenging, as previously discussed.  
We were, in effect, ``building the plane while flying it'': our technical infrastructure had to co-evolve with the project’s ideas, yet that same infrastructure was also required to collect artifacts and develop traces.  
This dynamic limited our ability to fully exploit the benefits of ongoing reflection during the project itself.  

Nevertheless, we see our experience as an important first step toward realizing these benefits.  
Traceable research practices may ultimately enhance research outcomes, much as early evidence suggests that open science practices positively influence research quality~\cite{soderberg_initial_2021}.  
By developing and sharing tools, workflows, and lessons learned, we aim to help others more easily test and extend these ideas.  
Demonstrating such benefits conclusively, however, will require coordinated and systematic investigation by the broader research community.  

\subsection{The Burdens of Tracing}
Over the past decade, the procedural demands on researchers have grown substantially.  
Increasingly explicit and implicit rules are intended to enhance fairness, equity, inclusion, and rigor.  
Examples include pre-registration, detailed supplementary materials with documented analysis scripts, supplementary videos, and accessibility-focused text descriptions.  
While these practices serve worthy goals, they also create tangible downsides, including the privileging of well-resourced teams able to meet requirements, increasing the workload of junior researchers,
and potentially slowing the pace of research.

Traceability exemplifies this tension.  
Documenting, reflecting on, and curating research processes \textit{can} improve rigor, transparency, and opportunities for others to learn.  
Deep linking to supplemental material \textit{can} enrich the reading experience.  
Yet these benefits come at a cost.  
Recording, digitizing, annotating, and organizing research artifacts requires substantial time and expertise.  
Even with supportive tools—such as automatic transcription and summarization, these tasks are labor-intensive and potentially disruptive to research flow. Not to mention the increased work of reading and reviewing papers that come with significantly more information to parse.

Community buy-in could help, and we see two complementary paths forward.  
First, the community could jointly invest in infrastructure for tRRRacer-like traceability, following precedents from the open science movement (e.g., OSF.io provides preregistration, supplemental material hosting, and anonymization features). 
Alternatively, we could develop low-tech standards for traceability, such as aggregated artifacts in supplementary PDFs, including thickly annotated threads, that are then included in reports via mechanisms such ase deep links.

However, open questions remain about the feasibility of tracing that we argue warrant further research. Is it possible to capture the tacit knowledge of researchers within artifacts? What kinds of artifacts should we be collecting? How much evidence and description is enough? What are appropriate incentives to encourage researchers to be more transparent? And ultimately, what is a reasonable balance between the benefits of rigor and transparency with the practical realities of research practice?

\subsection{\rev{Limitations}}
Our reflections on traceability are grounded in a design experiment conducted within our own research group, and therefore inherit the scope and biases of that setting. Furthermore, our experiment opens more questions than answers around how to support researchers and readers in tracing. Our work did not thoroughly explore all the kinds of artifacts that could be recorded, the ways artifacts may be annotated, or the myriad of opportunities for visualizing and communicating threads.

We did not evaluate the researcher or reader experience of tRRRacer with audiences beyond our team. 
As a result, we do not yet claim that threads and artifact collections are appropriately scoped or beneficial for broader audiences in realistic reading contexts. Our tRRRacer tool was an experiment, conducted in the designerly approach of \textit{research through design}; as such, we do not claim the tool as a final endpoint, or as an artifact worthy of extensive user testing. The development of tRRRacer was a sandbox for us, as a research team, to explore ideas and learn along the way. We see the potential for future studies to explore how threads and deep links, for example, can beneficially supplement the reading experience, as well as the testing of new tools and workflows for supporting bespoke and light-weight artifact collection and curation.

\section{Conclusions}
We introduced traceability as a framework for making design-oriented visualization research more transparent and scrutinizable.  
Through our tRRRacer design experiment, we explored how recording, reporting, and reading can work together to surface the origins and development of research contributions.  
Our reflections revealed both the promise and the burden of implementing traceability: it enriches rigor, transparency, and learning, yet requires time, infrastructure, and thoughtful curation.  

We offer our theoretical framing of traceability as both a generative tool for thoughtful recording and reporting of visualization research, as well as a goal for the community to consider from a socio-technical perspective. We hope this paper seeds discussions towards concrete next steps for lightweight practices and community standards for traceability that can support our increasingly diverse and pluralistic research community.

\section{Acknowledgments}
This work is funded in part by the Swedish Research Council (Grant No. 2024-05726); by the Wallenberg
AI, Autonomous Systems and Software Program (WASP) funded by
the Knut and Alice Wallenberg Foundation; and the U.S. National Science Foundation (NSF OAC 1835904).

\bibliographystyle{eg-alpha-doi} 
\bibliography{vr_proposal}       

\end{document}